\documentclass[%
 reprint,
 showkeys,
superscriptaddress,
amsmath,amssymb,
aps,
pra,
]{revtex4-2}

\usepackage{float}
\usepackage{mathrsfs}
\usepackage{graphicx}
\usepackage{physics}
\usepackage{hyperref}
\usepackage{cleveref}
\usepackage{bbm}
\usepackage{booktabs}%
\usepackage[dvipsnames]{xcolor}
\hypersetup{
        colorlinks   = true,
        citecolor    = LimeGreen,
        linkcolor    = LimeGreen}

\begin{document}
\preprint{APS/123-QED}

\title{Dynamical and Emission Properties of Quantum \\ Emitters driven by Ultra-Short Laser Pulses}


\author{Juan E. Ardila-García}
\email{juardilag@unal.edu.co}
\affiliation{Grupo de Superconductividad y Nanotecnología, Departamento de Física, Universidad Nacional de Colombia, Bogotá, 111321, Colombia}

\author{Juan S. Sierra-Jaraba}
\affiliation{Grupo de Superconductividad y Nanotecnología, Departamento de Física, Universidad Nacional de Colombia, Bogotá, 111321, Colombia}

\author{Herbert Vinck-Posada}
\affiliation{Grupo de Superconductividad y Nanotecnología, Departamento de Física, Universidad Nacional de Colombia, Bogotá, 111321, Colombia}

\date{\today}

\begin{abstract}
The development of high-performance quantum technologies relies on the ability to prepare the quantum states of solid-state emitters with high fidelity while cleanly separating the emitted photons from the driving field. Here, we present a comprehensive theoretical comparison of three single-pulse coherent control protocols: resonant Rabi oscillations, adiabatic rapid passage (ARP), and notch-filtered ARP (NARP). To establish an ideal performance baseline, we first map the parameter spaces for each protocol in a closed system, identifying the regions of robust population inversion. We then use a Lindblad master equation to compute the time-resolved emission spectra in the presence of decoherence. Our results show that while all three schemes can generate identical, transform-limited Lorentzian photons, NARP uniquely combines the high-fidelity robustness of adiabatic passage with intrinsic spectral separability. Our findings, which align with the work on NARP by Wilbur et al. \cite{Wilbur2022APL}, establish a clear design framework for engineering the next generation of quantum light sources.
\end{abstract}

\maketitle

\section{Introduction}

Ultrafast optical control of two-level quantum systems (TLSs) is a foundational capability for developing scalable solid-state quantum technologies~\cite{Shields2007}. Among the most promising platforms are semiconductor quantum dots (QDs), which function as ``artificial atoms" offering discrete energy levels, strong dipole transitions, and seamless integration with photonic nanostructures~\cite{Warburton2013,Senellart2017}. These attributes have enabled the development of high-brightness and near-transform-limited single-photon sources~\cite{Kuhlmann2015,Somaschi2016,Tomm2021}, which are critical components for quantum communication protocols~\cite{AspuruGuzik2012} and spin-based quantum computing architectures~\cite{Loss1998}.

A central challenge in the coherent control of QDs is achieving high-fidelity state preparation while effectively suppressing scattered light from the driving laser. In resonant excitation schemes, the emitted photons are at the same frequency as the laser, necessitating stringent filtering techniques—such as polarization filtering or interferometric suppression—to isolate the quantum signal~\cite{Kuhlmann2013,Somaschi2016}. However, such methods are often limited in their extinction ratio and inherently reduce collection efficiency. Alternative non-resonant approaches, including off-resonant excitation or phonon-assisted processes, can circumvent direct spectral overlap but often at the cost of reduced photon indistinguishability or added experimental complexity~\cite{Ramsay2010phonons,IlesSmith2017}.

Ultrafast pulse shaping provides a powerful alternative for robust state preparation and clean spectral separation~\cite{Vandersypen2005}. Coherent population inversion via Rabi oscillations, induced by transform-limited pulses, is a fundamental technique but remains highly sensitive to fluctuations in envelope area and detuning~\cite{Allen1987,Press2008}. In contrast, adiabatic rapid passage (ARP), which employs frequency-chirped pulses, enables robust, broadband inversion by adiabatically following the dressed eigenstates of the system~\cite{Malinovsky2001,Vitanov2017}. Experimental implementations of ARP in QDs have demonstrated high-fidelity inversion with excellent photon indistinguishability~\cite{Wu2011,Wei2014,Simon2011}. More recently, notch-filtered ARP (NARP) was introduced to combine the robustness of ARP with intrinsic spectral filtering~\cite{Wilbur2022APL}. By creating a narrow spectral notch at the TLS resonance, NARP aims to suppress the spectral overlap between the driving pulse and the emitted fluorescence, thereby enabling background-free detection without compromising inversion fidelity. It has been shown that both the ARP and NARP pulse shapes can be achieved using amplitude and phase masks. \cite{Wilbur2022APL}

\begin{figure}[H]
    \centering
    \includegraphics[scale = 1]{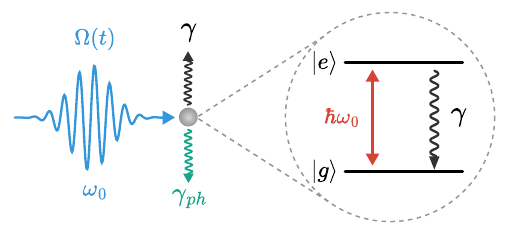}
    \caption{Model schematic: a two-level system (TLS) with transition energy $\hbar\omega_0$ is driven by a single resonant shaped ultra-short laser pulse with Rabi frequency $\Omega(t)$, and subject to spontaneous emission ($\gamma$) and pure dephasing ($\gamma_{ph}$).}
    \label{fig:model}
\end{figure}

From a theoretical standpoint, the emission from a coherently driven TLS is fundamentally shaped by its interaction with the environment and the spectro-temporal properties of the excitation field~\cite{Scully1997,Breuer2002}. For strong resonant driving, the emission spectrum famously splits into a Mollow triplet~\cite{Mollow1969}. However, under pulsed excitation, the emission characteristics depend critically on the pulse shape and duration. When environmental correlations are short-lived (the Markovian approximation), the system dynamics are well described by a master equation in Lindblad form, which accounts for both radiative decay and pure dephasing~\cite{Carmichael1999,Krummheuer2002,Manzano2020}. For quantum dots, a dephasing time in the range of a few nanoseconds is mentioned as the ultimate limit for the long exponential decay of optically-driven excitation, which is eventually limited by the radiative lifetime \cite{Borri2001}. Specifically, research on epitaxially-grown InGaAs/GaAs quantum dots at low temperatures observed a long exponential decay (several hundreds of picoseconds), which was ultimately limited by the radiative lifetime, typically in the few nanosecond range \cite{Borri2001}. In this regime, the emission from a fully inverted TLS is expected to be a single Lorentzian peak, broadened by decoherence mechanisms~\cite{Shore2011}. While the state-preparation fidelities of Rabi, ARP, and NARP schemes are well-documented~\cite{Ramsay2010, Malinovsky2001,Stark2022,Wilbur2022APL}, a systematic comparison of their resulting emission spectra and their robustness against parameter noise in the context of spectral filtering is still needed.

In this work, we present a comprehensive theoretical comparison of the Rabi, ARP, and NARP control protocols for single-pulse excitation of a QD-based TLS. Using a Lindblad master equation that includes both radiative and pure-dephasing channels, we systematically evaluate the time-domain dynamics, parameter robustness, and time-dependent emission spectra for each scheme. We demonstrate that while all three protocols can generate indistinguishable, Lorentzian-shaped photons under ideal conditions, NARP uniquely facilitates the spectral separation of fluorescence from the a driving pulse. By mapping the control parameter spaces—envelope area, chirp rate, and notch width—we identify the regions that ensure robust population inversion in a closed system. These maps provide a crucial baseline for understanding performance limits and guiding experimental implementations where decoherence is weak but non-negligible. Our results establish a practical framework for selecting and engineering excitation protocols for high-performance quantum light sources.

\section{Theoretical Model}
\label{sec:theory}

We model the system as a two-level quantum system (TLS) with a ground state $|g\rangle$ and an excited state $|e\rangle$, separated by an energy $\hbar\omega_0$. This model serves as an effective description of a single exciton transition in a semiconductor quantum dot. As depicted in Fig.~\ref{fig:model}, the TLS is driven by a classical laser field and coupled to a reservoir that induces spontaneous emission and pure dephasing.

The driving laser pulse is described by a classical electric field $\mathbf{E}(t)$ within the dipole approximation:
\begin{equation}
    \mathbf{E}(t) = \boldsymbol{\epsilon} E_0(t) \cos[\omega_L t + \varphi(t)],
\end{equation}
where $\boldsymbol{\epsilon}$ is the polarization unit vector, $E_0(t)$ is the real pulse envelope, $\omega_L$ is the laser carrier frequency, and $\varphi(t)$ is a time-dependent phase that defines the pulse's frequency chirp.

In the laboratory frame and under the rotating-wave approximation (RWA)~\cite{Allen1987}, the system Hamiltonian is
\begin{equation}
    H(t) = -\frac{\hbar \omega_0}{2} \sigma_z - \frac{\hbar}{2} \left[ \Omega(t) e^{-i \omega_L t} \sigma_+ + \Omega^*(t) e^{i \omega_L t} \sigma_- \right],
\end{equation}
where $\sigma_z = |e\rangle\langle e| - |g\rangle\langle g|$, $\sigma_+ = |e\rangle\langle g|$, and $\sigma_- = |g\rangle\langle e|$ are the standard Pauli operators. The complex Rabi frequency is defined as $\Omega(t) = \hbar^{-1}\mathbf{d}_{eg} \cdot \boldsymbol{\epsilon} E_0(t) e^{-i \varphi(t)}$, with $\mathbf{d}_{eg}$ being the transition dipole moment.

To remove the fast optical oscillations, we transform into a rotating frame via the unitary operator $U(t) = \exp[i\omega_L t \sigma_z/2]$. The Hamiltonian in this frame becomes
\begin{equation}
    \tilde{H}(t) = -\frac{\hbar}{2} (\omega_0 - \omega_L) \sigma_z - \frac{\hbar}{2} \left[ \Omega(t) \sigma_+ + \Omega^*(t) \sigma_- \right].
\end{equation}
For a chirped pulse (ARP-NARP), it is convenient to move into a second generalized rotating frame that follows the pulse phase, yielding the final Hamiltonian~\cite{Shore2011}
\begin{equation}
    H'(t) = -\frac{\hbar}{2} \Delta(t) \sigma_z - \frac{\hbar}{2} |\Omega(t)| \sigma_x,
    \label{eq:final_hamiltonian}
\end{equation}
where the instantaneous detuning is $\Delta(t) = (\omega_0 - \omega_L) - \dot{\varphi}(t)$. This Hamiltonian is of the Landau-Zener form, and its dynamics are adiabatic when the rate of change of the Hamiltonian's direction in Hilbert space is much smaller than the energy gap between its instantaneous eigenstates~\cite{Landau1932, Zener1932, Shevchenko2010}. This leads to the well-known adiabaticity condition~\cite{Malinovsky2001}:
\begin{equation}
    \frac{\left| \dot{\Delta}(t)|\Omega(t)| - \Delta(t)|\dot{\Omega}(t)| \right|}{\left[ |\Omega(t)|^2 + \Delta(t)^2 \right]^{3/2}} \ll 1.
    \label{eq:adiabatic_condition}
\end{equation}

To account for incoherent processes, we model the evolution of the system using the Lindblad master equation for the density matrix $\rho(t)$:
\begin{equation}
    \frac{d\rho(t)}{dt} = -\frac{i}{\hbar} [H(t), \rho(t)] + \mathcal{L}_{\mathbf{\sigma_-}}[\rho(t)] + \mathcal{L}_{\mathbf{\sigma}_z}[\rho(t)],
\end{equation}
where the Lindblad superoperators $\mathcal{L}_{\mathbf{\sigma_-}}$ and $\mathcal{L}_{\mathbf{\sigma}_z}$ describe spontaneous emission and pure dephasing, respectively. The spontaneous emission channel is given by
\begin{equation}
    \mathcal{L}_{\mathbf{\sigma_-}}[\rho] = \frac{\gamma}{2} \left(2 \sigma_- \rho \sigma_+ - \sigma_+ \sigma_- \rho - \rho \sigma_+ \sigma_-\right),
\end{equation}
where $\gamma$ is the radiative decay rate. Pure dephasing, which in QDs arises primarily from elastic scattering with acoustic phonons, is modeled by~\cite{Borri2001, Krummheuer2002}
\begin{equation}
    \mathcal{L}_{\mathbf{\sigma}_z}[\rho] = \gamma_{\mathrm{ph}} \left(\sigma_z \rho \sigma_z - \rho\right).
\end{equation}
Here, $\gamma_{\mathrm{ph}}$ is the phenomenological pure dephasing rate.

The emission spectrum of the TLS is, except for a constant, given by the Fourier transform of the first-order photon correlation function, $g^{(1)}(t, t+\tau) = \langle \sigma_+(t+\tau) \sigma_-(t) \rangle$. For pulsed excitation, we compute the time-dependent physical spectrum (spectrogram)~\cite{Eberly1977}:
\begin{equation}
    S(\omega, t) \propto \mathrm{Re} \int_0^\infty d\tau\, e^{-i\omega \tau} \langle \sigma_+(t+\tau) \sigma_-(t) \rangle.
\end{equation}
The two-time correlation function is calculated by evolving the system to time $t$, applying the lowering operator $\sigma_-(t)$ to the density matrix, and then evolving the resulting operator forward in time according to the quantum regression theorem~\cite{Lax1963, Carmichael1999, Swain1981}.

\section{Coherent Control Protocols}
\label{sec:protocols}

The interaction between a TLS and a spectrally shaped ultrafast pulse gives rise to distinct dynamical regimes. Here, we analyze three canonical protocols, from the simplest one: transform-limited Rabi oscillations to more engineered mechanisms: adiabatic rapid passage (ARP) using chirped pulses, and the recently developed notch-filtered ARP (NARP)~\cite{Wilbur2022APL,Stark2022}. These protocols represent a progression from fast, resonant control toward robust, error-tolerant, and spectrally filterable population transfer and also the three of them in order: Rabi oscillations, ARP and NARP represent a logical progression from resonant light use for population transfer to intentionally shaped pulses engineered towards the improvement of excitation properties, robustness, and spectral separability. For a direct comparison, all protocols are based on pulses with an underlying Gaussian spectral envelope centered at the TLS resonance, $\omega_L = \omega_0$. The specific pulse shapes are generated by applying phase and amplitude masks in the frequency domain before inverse Fourier transformation.    

\begin{figure*}[t]
    \centering
    \includegraphics[scale = 1]{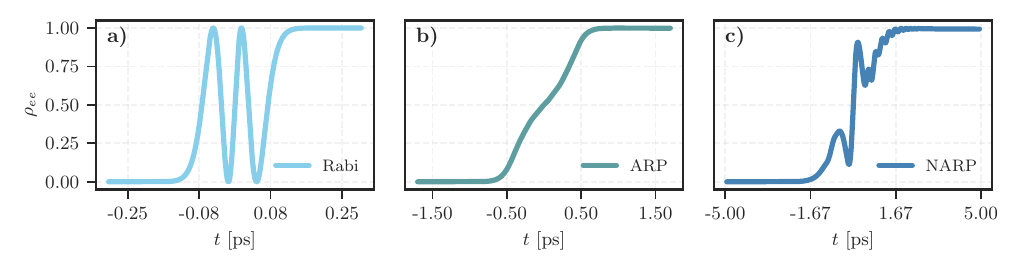}
    \caption{Time evolution of the excited-state population, $\rho_{ee}(t)$, for the three excitation protocols. Each pulse has a total area of $\Theta=5\pi$ and a transform-limited temporal intensity FWHM of $\tau_0 = 100$~fs (corresponding to $\tau_0 \approx 0.44/\Gamma_0$).
    (\textbf{a}) \textit{Rabi oscillations}: A transform-limited pulse induces coherent population cycling.
    (\textbf{b}) \textit{Adiabatic rapid passage (ARP)}: A linearly chirped pulse ($\alpha/\tau_0^2 = 0.8$) produces smooth and complete population inversion, though the pulse is temporally broadened.
    (\textbf{c}) \textit{Notch-filtered ARP (NARP)}: A chirped pulse ($\alpha/\tau_0^2 = 2.4$) with a Gaussian spectral notch ($\delta/\Gamma_0 = 0.25$) also achieves adiabatic inversion, but the dynamics are modulated by fast oscillations arising from temporal interference effects.}
    \label{fig:excitation_schemes}
\end{figure*}

\subsection{Resonant Rabi Oscillations}

\begin{figure}[t]
    \centering
    \includegraphics[scale = 1]{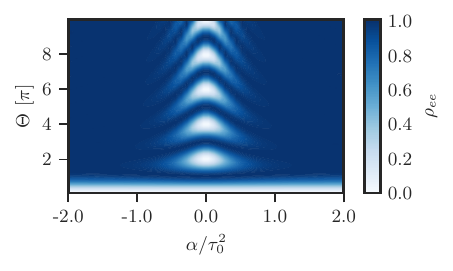}
    \caption{Final excited-state population $\rho_{ee}$ for ARP as a function of envelope area $\Theta$ and normalized chirp $\alpha/\tau_0^2$. The vertical slice at $\alpha = 0$ corresponds to the resonant Rabi regime. For increasing chirp ($|\alpha/\tau_0^2| \gtrsim 1.5$), a robust plateau of near-unity inversion emerges for all envelope areas $\Theta \gtrsim \pi$, demonstrating the protocol's tolerance to laser intensity fluctuations.}
    \label{fig:arp_area_scan}
\end{figure}

The most direct excitation method uses a transform-limited Gaussian pulse with a flat spectral phase, resonant with the TLS transition. In this case, the instantaneous detuning is zero, $\Delta(t)=0$, and the Hamiltonian in the rotating frame simplifies to
\begin{equation}
    H'(t) = -\frac{\hbar}{2}|\Omega(t)|\sigma_x.
\end{equation}
The evolution is governed by the pulse-area theorem, which dictates that the final excited-state population is a sinusoidal function of the total Rabi frequency envelope area $\Theta = \int_{-\infty}^{\infty} |\Omega(t)| dt$~\cite{Allen1987}:
\begin{equation}
    \rho_{ee}(\infty) = \sin^2\left(\frac{\Theta}{2}\right).
\end{equation}
As shown in Fig.~\ref{fig:excitation_schemes}(a), this protocol enables ultrafast population cycling. However, its practical application is limited by a high sensitivity to fluctuations in envelope area and frequency, as illustrated by the sharp, oscillatory features along the $\alpha=0$ axis in Fig.~\ref{fig:arp_area_scan}.

\subsection{Adiabatic Rapid Passage (ARP)}

ARP achieves robust population inversion by using a quadratic spectral phase, $\phi(\omega) = \frac{\alpha}{2}(\omega - \omega_L)^2$, which creates a linear frequency sweep in the time domain. This results in a time-dependent detuning $\Delta(t)$ that sweeps through the resonance. Provided the evolution is slow compared to the energy gap between the instantaneous dressed states, the system adiabatically follows one eigenstate from $|g\rangle$ to $|e\rangle$, resulting in complete inversion~\cite{Vitanov2017}. This process is robust as long as the adiabaticity condition [Eq.~\eqref{eq:adiabatic_condition}] is met.

Figure~\ref{fig:excitation_schemes}(b) shows the smooth population transfer characteristic of ARP. The protocol's key advantage—robustness—is mapped in Fig.~\ref{fig:arp_area_scan} (closed system). For small chirp, the dynamics are dominated by non-adiabatic Rabi-like oscillations. As the chirp increases to $|\alpha/\tau_0^2| \gtrsim 1.5$, a broad, stable plateau of near-unity inversion ($\rho_{ee} \approx 1$) develops, indicating that the final state becomes largely insensitive to the precise envelope area, provided it is above a threshold value (here, $\Theta \gtrsim \pi$).

\begin{figure*}[t]
    \centering
    \includegraphics[width=\textwidth]{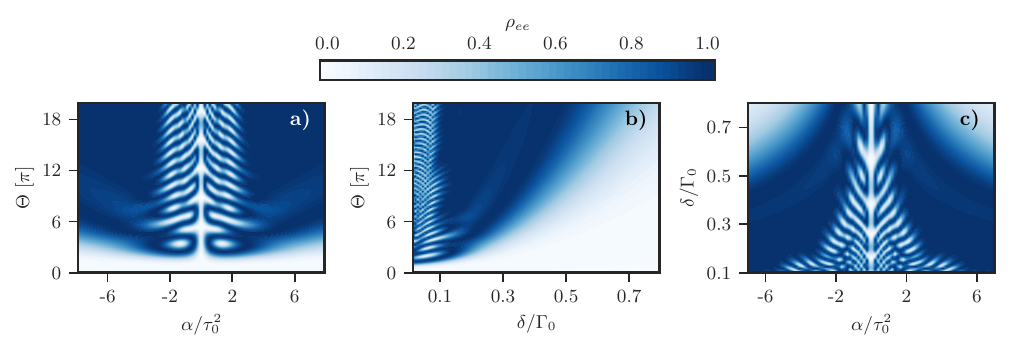}
    \caption{Parameter space for the final excited-state population $\rho_{ee}$ under NARP excitation, shown as 2D slices through the $(\Theta, \alpha, \delta)$ parameter volume.
    (\textbf{a}) $\rho_{ee}$ vs. envelope area $\Theta$ and chirp $\alpha$ for a fixed notch width $\delta/\Gamma_0 = 0.25$. A robust inversion plateau appears for large chirp and area.
    (\textbf{b}) $\rho_{ee}$ vs. envelope area $\Theta$ and notch width $\delta$ for a fixed chirp $\alpha/\tau_0^2 = 5$. Wider notches require a larger envelope area to achieve inversion.
    (\textbf{c}) $\rho_{ee}$ vs. chirp $\alpha$ and notch width $\delta$ for a fixed envelope area $\Theta = 16\pi$. Maintaining inversion with wider notches requires a corresponding increase in chirp.
    The oscillatory patterns are signatures of Landau-Zener-St{\"u}ckelberg interference from non-adiabatic transitions near the avoided crossing.}
    \label{fig:narp_combined_scan}
\end{figure*}

\subsection{Notch-Filtered ARP (NARP)}

While robust, ARP uses a pulse whose spectrum significantly overlaps with the TLS transition, complicating the filtering of emitted photons. NARP addresses this by engineering a spectral hole at the transition frequency. The pulse in the frequency domain is described by
\begin{equation}
    \tilde{\Omega}(\omega) = \tilde{\Omega}_{\text{Gauss}}(\omega) \left[1 - e^{-(\omega-\omega_0)^2/(2\delta^2)}\right] e^{i\frac{\alpha}{2}(\omega-\omega_L)^2},
\end{equation}
where $\tilde{\Omega}_{\text{Gauss}}(\omega)$ is the underlying Gaussian spectrum and $\delta$ is the width of the Gaussian notch. The removal of resonant frequency components creates a more complex temporal pulse shape, as seen in the rapid oscillations superimposed on the population dynamics in Fig.~\ref{fig:excitation_schemes}(c).

Despite the absence of resonant driving fields, adiabatic inversion is still possible~\cite{Wilbur2022APL,Stark2022}. Figure~\ref{fig:narp_combined_scan} maps the operational parameter space for NARP (closed system). The three panels confirm that robust inversion plateaus ($\rho_{ee} \approx 1$) persist, governed by clear trade-offs:
(\textbf{a}) Similar to ARP, a large chirp and envelope area are required.
(\textbf{b}) At a fixed chirp, increasing the notch width $\delta$ necessitates a larger envelope area $\Theta$ to compensate for the removed energy.
(\textbf{c}) At a fixed envelope area, a wider notch must be paired with a stronger chirp to maintain adiabaticity.
The intricate fringe patterns visible in all panels are characteristic of Landau-Zener-St{\"u}ckelberg interference, arising from non-adiabatic coupling between the dressed states~\cite{Shevchenko2010}. These maps demonstrate that NARP retains the robustness of ARP while enabling spectral filtering, provided the parameters are chosen within the identified high-fidelity regions.

\section{Emission Spectrum and Practical Implications}
\label{sec:emission}

The utility of a quantum light source is ultimately determined by the properties of the photons it emits. For applications requiring high-purity, background-free single photons, the ability to spectrally separate the emitted signal from the driving laser is paramount. To analyze this, we compute the time-dependent physical spectrum (spectrogram) for each protocol, which is derived from the two-time correlation function $\langle\sigma_+(t+\tau)\sigma_-(t)\rangle$ via the quantum regression theorem~\cite{Carmichael1999, Eberly1977}.

\begin{figure*}[t]
    \centering
    \includegraphics[width=\textwidth]{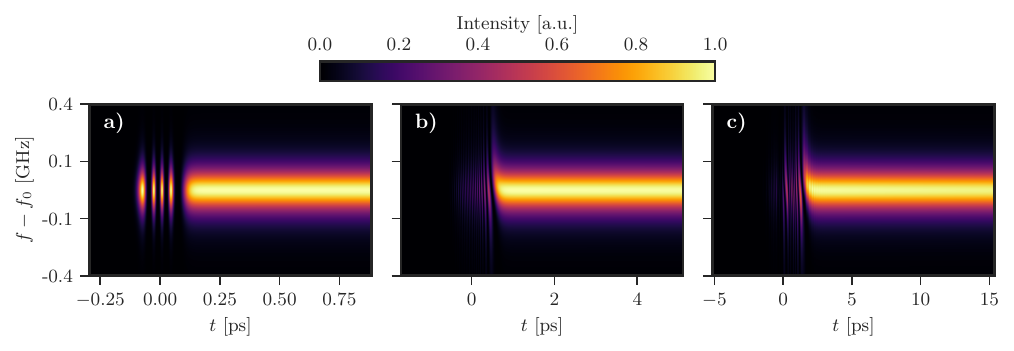}
    \caption{Time-resolved emission spectrograms for (\textbf{a}) Rabi ($\Theta=9\pi$), (\textbf{b}) ARP ($\Theta=9\pi$, $\alpha/\tau_0^2=0.4$), and (\textbf{c}) NARP ($\Theta=9\pi$, $\alpha/\tau_0^2=1.2$, $\delta/\gamma=0.25$) excitation. While all three generate an identical Lorentzian emission line at the TLS frequency $f_0$, their temporal characteristics differ. Rabi excitation shows emission during the pulse, followed by strong free-induction decay post-excitation ($t>0.1$~ps). In contrast, the positive chirp in ARP and NARP delays the emission onset. This effect is most pronounced for NARP, where the combination of a large chirp and the spectral notch pushes the start of significant emission to several picoseconds after the pulse center.}
    \label{fig:emission_spectrograms}
\end{figure*}

Figure~\ref{fig:emission_spectrograms} presents the calculated spectrograms for the three control schemes, using parameters representative of high-quality InAs/GaAs QDs at cryogenic temperatures: a radiative lifetime $T_1 = 1/\gamma = 1$~ns and a pure dephasing time $T_2^* = 1/\gamma_{\mathrm{ph}} = 10$~ns~\cite{Kuhlmann2015}. The central result is that, despite their vastly different underlying dynamics, all three protocols produce an identical emission spectrum after the pulse has passed: a clean Lorentzian line centered precisely at the transition frequency $f_0$. The full width at half maximum (FWHM) of this line is given by the total decoherence rate $\Gamma = \gamma + \gamma_{\mathrm{ph}}$ (here, $\approx 175$~MHz), confirming that the emitted photons are transform-limited by the emitter's coherence, irrespective of the excitation method.

While the final emission is identical, the temporal dynamics during excitation reveal crucial differences. As seen in Fig.~\ref{fig:emission_spectrograms}(a), Rabi driving produces transient spectral features during the pulse, corresponding to the population cycling. In contrast, the chirped pulses of ARP and NARP delay the onset of emission until the instantaneous frequency sweeps into resonance [Figs.~\ref{fig:emission_spectrograms}(b) and (c)]. This temporal separation is a direct consequence of the adiabatic inversion mechanism.

These findings have significant practical consequences for spectral filtering. The key advantage of NARP is that it creates a "dark" window at the emitter's frequency in the excitation spectrum. For a pulse with a transform-limited duration of $\tau_0 \approx 100$~fs, the spectral bandwidth is on the order of terahertz. A notch width of $\delta$ corresponding to a few GHz is therefore orders of magnitude wider than the sub-GHz emission linewidth but still represents a tiny fraction of the total pulse bandwidth. This vast separation in scales ensures that a simple, fixed-frequency spectral filter (e.g., a Fabry-Pérot etalon) can completely reject the scattered laser light while transmitting the entire photon wavepacket, making NARP uniquely suited for generating high-purity photons.

Finally, the performance of these protocols is intrinsically linked to the decoherence environment. Enhancing the radiative rate $\gamma$ via cavity or nanostructure coupling (i.e., the Purcell effect)~\cite{Claudon2010, Arcari2014} broadens the emission linewidth, which relaxes filtering requirements but demands faster control pulses to complete inversion before decay occurs. Conversely, pure dephasing from phonon scattering~\cite{Ramsay2010phonons, Vagov2007} limits the maximum allowable pulse duration and chirp. An excessively long or heavily chirped pulse will interact with the system for too long, allowing dephasing to degrade the coherence and reduce the final inversion fidelity. This establishes a fundamental trade-off between robustness, spectral filterability, and the intrinsic coherence properties of the emitter.

\section{Discussion}

Our analysis provides a systematic comparison of three distinct coherent control protocols by mapping their ideal parameter landscapes in a closed system (Figs.~\ref{fig:arp_area_scan} and \ref{fig:narp_combined_scan}) and then simulating their dynamics and emission in an open system (Figs.~\ref{fig:excitation_schemes} and \ref{fig:emission_spectrograms}). The closed-system maps reveal the intrinsic robustness of each protocol to parameter fluctuations. While resonant Rabi driving is highly sensitive, ARP introduces significant robustness by facilitating an adiabatic evolution of the system's dressed states~\cite{Malinovsky2001,Vitanov2017}.

The NARP protocol, introduced by Wilbur \textit{et al.}~\cite{Wilbur2022APL}, extends this robustness while enabling spectral filtering. Our closed-system maps corroborate their findings, showing that robust adiabatic inversion is achievable despite the absence of resonant driving components. The open-system simulations further show that this high-fidelity control translates directly into the generation of clean, transform-limited photons. The essential design principle is that the absolute width of the spectral notch must be significantly larger than the emitter's decoherence-broadened linewidth ($\delta \gg \Gamma$), a condition that is easily met in practice. Our findings are in excellent agreement with the original theoretical and experimental work on NARP~\cite{Wilbur2022APL, Stark2022}, confirming its viability for high-performance quantum light sources.

\section{Conclusion}

In summary, we have conducted a comprehensive theoretical comparison of Rabi, ARP, and NARP excitation protocols. By first analyzing the ideal, closed-system parameter space, we have delineated the intrinsic robustness of each method. Subsequent open-system simulations confirm that while all three schemes can generate transform-limited photons, they offer vastly different trade-offs between speed, robustness, and spectral filterability.

Rabi driving offers speed at the cost of precision. ARP provides robustness but no intrinsic filtering. NARP, as conceived by Wilbur \textit{et al.}~\cite{Wilbur2022APL}, successfully combines the robustness of adiabatic passage with the practical necessity of spectral filtering. Our theoretical framework, which corroborates their experimental demonstrations, establishes NARP as a superior and practical control scheme for generating high-purity single photons from solid-state emitters. This work provides quantitative guidance for the engineering of next-generation quantum photonic devices, where the interplay between ideal control landscapes and real-world decoherence must be carefully managed.

\section*{Acknowledgements}

J. E. A.-G. thanks the Quantum Nanophotonics Lab at Queen’s University, Canada, for its support during a summer research internship, which was essential for the initial development of this work. Special acknowledgment is given to Nir Rotenberg and Tristan Austin for their guidance, and to Sofia Arranz Regidor for her valuable discussions on the theoretical analysis of the emission spectrum. All authors acknowledge funding from the project ``Ampliación del
uso de la mecánica cuántica desde el punto de vista experimental y su relación con la teoría, generando desarrollos en tecnologías cuánticas útiles para metrología y computación cuántica a nivel nacional", BPIN
2022000100133, from SGR of MINCIENCIAS, Gobierno de Colombia

\bibliography{main_bib}

\begin{thebibliography}{39}%
\makeatletter
\providecommand \@ifxundefined [1]{%
 \@ifx{#1\undefined}
}%
\providecommand \@ifnum [1]{%
 \ifnum #1\expandafter \@firstoftwo
 \else \expandafter \@secondoftwo
 \fi
}%
\providecommand \@ifx [1]{%
 \ifx #1\expandafter \@firstoftwo
 \else \expandafter \@secondoftwo
 \fi
}%
\providecommand \natexlab [1]{#1}%
\providecommand \enquote  [1]{``#1''}%
\providecommand \bibnamefont  [1]{#1}%
\providecommand \bibfnamefont [1]{#1}%
\providecommand \citenamefont [1]{#1}%
\providecommand \href@noop [0]{\@secondoftwo}%
\providecommand \href [0]{\begingroup \@sanitize@url \@href}%
\providecommand \@href[1]{\@@startlink{#1}\@@href}%
\providecommand \@@href[1]{\endgroup#1\@@endlink}%
\providecommand \@sanitize@url [0]{\catcode `\\12\catcode `\$12\catcode `\&12\catcode `\#12\catcode `\^12\catcode `\_12\catcode `\%12\relax}%
\providecommand \@@startlink[1]{}%
\providecommand \@@endlink[0]{}%
\providecommand \url  [0]{\begingroup\@sanitize@url \@url }%
\providecommand \@url [1]{\endgroup\@href {#1}{\urlprefix }}%
\providecommand \urlprefix  [0]{URL }%
\providecommand \Eprint [0]{\href }%
\providecommand \doibase [0]{https://doi.org/}%
\providecommand \selectlanguage [0]{\@gobble}%
\providecommand \bibinfo  [0]{\@secondoftwo}%
\providecommand \bibfield  [0]{\@secondoftwo}%
\providecommand \translation [1]{[#1]}%
\providecommand \BibitemOpen [0]{}%
\providecommand \bibitemStop [0]{}%
\providecommand \bibitemNoStop [0]{.\EOS\space}%
\providecommand \EOS [0]{\spacefactor3000\relax}%
\providecommand \BibitemShut  [1]{\csname bibitem#1\endcsname}%
\let\auto@bib@innerbib\@empty
\bibitem [{\citenamefont {Wilbur}\ \emph {et~al.}(2022)\citenamefont {Wilbur}, \citenamefont {Meesala}, \citenamefont {Son}, \citenamefont {Allen}, \citenamefont {Lu}, \citenamefont {Bersin}, \citenamefont {Mouradian}, \citenamefont {Stark},\ and\ \citenamefont {Englund}}]{Wilbur2022APL}%
  \BibitemOpen
  \bibfield  {author} {\bibinfo {author} {\bibfnamefont {C.~C.}\ \bibnamefont {Wilbur}}, \bibinfo {author} {\bibfnamefont {S.}~\bibnamefont {Meesala}}, \bibinfo {author} {\bibfnamefont {G.}~\bibnamefont {Son}}, \bibinfo {author} {\bibfnamefont {J.~P.}\ \bibnamefont {Allen}}, \bibinfo {author} {\bibfnamefont {T.-H.}\ \bibnamefont {Lu}}, \bibinfo {author} {\bibfnamefont {E.}~\bibnamefont {Bersin}}, \bibinfo {author} {\bibfnamefont {S.~L.}\ \bibnamefont {Mouradian}}, \bibinfo {author} {\bibfnamefont {C.~J.}\ \bibnamefont {Stark}},\ and\ \bibinfo {author} {\bibfnamefont {D.}~\bibnamefont {Englund}},\ }\bibfield  {title} {\bibinfo {title} {Notch-filtered adiabatic rapid passage for optically driven quantum light sources},\ }\href {https://doi.org/10.1063/5.0090048} {\bibfield  {journal} {\bibinfo  {journal} {APL Photonics}\ }\textbf {\bibinfo {volume} {7}},\ \bibinfo {pages} {111302} (\bibinfo {year} {2022})}\BibitemShut {NoStop}%
\bibitem [{\citenamefont {Shields}(2007)}]{Shields2007}%
  \BibitemOpen
  \bibfield  {author} {\bibinfo {author} {\bibfnamefont {A.~J.}\ \bibnamefont {Shields}},\ }\bibfield  {title} {\bibinfo {title} {Semiconductor quantum dots as sources of single photons and entangled photon pairs},\ }\href {https://doi.org/10.1038/nphoton.2007.46} {\bibfield  {journal} {\bibinfo  {journal} {Nat. Photonics}\ }\textbf {\bibinfo {volume} {1}},\ \bibinfo {pages} {215} (\bibinfo {year} {2007})}\BibitemShut {NoStop}%
\bibitem [{\citenamefont {Warburton}(2013)}]{Warburton2013}%
  \BibitemOpen
  \bibfield  {author} {\bibinfo {author} {\bibfnamefont {R.~J.}\ \bibnamefont {Warburton}},\ }\bibfield  {title} {\bibinfo {title} {Single-photon sources in charge-tunable quantum dots},\ }\href {https://doi.org/10.1038/nmat3585} {\bibfield  {journal} {\bibinfo  {journal} {Nat. Mater.}\ }\textbf {\bibinfo {volume} {12}},\ \bibinfo {pages} {483} (\bibinfo {year} {2013})}\BibitemShut {NoStop}%
\bibitem [{\citenamefont {Senellart}\ \emph {et~al.}(2017)\citenamefont {Senellart}, \citenamefont {Solomon},\ and\ \citenamefont {White}}]{Senellart2017}%
  \BibitemOpen
  \bibfield  {author} {\bibinfo {author} {\bibfnamefont {P.}~\bibnamefont {Senellart}}, \bibinfo {author} {\bibfnamefont {G.}~\bibnamefont {Solomon}},\ and\ \bibinfo {author} {\bibfnamefont {A.}~\bibnamefont {White}},\ }\bibfield  {title} {\bibinfo {title} {High-performance semiconductor quantum-dot single-photon sources},\ }\href {https://doi.org/10.1038/nnano.2017.218} {\bibfield  {journal} {\bibinfo  {journal} {Nat. Nanotechnol.}\ }\textbf {\bibinfo {volume} {12}},\ \bibinfo {pages} {1026} (\bibinfo {year} {2017})}\BibitemShut {NoStop}%
\bibitem [{\citenamefont {Kuhlmann}\ \emph {et~al.}(2015)\citenamefont {Kuhlmann}, \citenamefont {Prechtel}, \citenamefont {Houel}, \citenamefont {Ludwig}, \citenamefont {Reuter}, \citenamefont {Wieck},\ and\ \citenamefont {Warburton}}]{Kuhlmann2015}%
  \BibitemOpen
  \bibfield  {author} {\bibinfo {author} {\bibfnamefont {A.~V.}\ \bibnamefont {Kuhlmann}}, \bibinfo {author} {\bibfnamefont {J.~H.}\ \bibnamefont {Prechtel}}, \bibinfo {author} {\bibfnamefont {J.}~\bibnamefont {Houel}}, \bibinfo {author} {\bibfnamefont {A.}~\bibnamefont {Ludwig}}, \bibinfo {author} {\bibfnamefont {D.}~\bibnamefont {Reuter}}, \bibinfo {author} {\bibfnamefont {A.~D.}\ \bibnamefont {Wieck}},\ and\ \bibinfo {author} {\bibfnamefont {R.~J.}\ \bibnamefont {Warburton}},\ }\bibfield  {title} {\bibinfo {title} {Transform-limited single photons from a single quantum dot},\ }\href {https://doi.org/10.1038/ncomms9204} {\bibfield  {journal} {\bibinfo  {journal} {Nat. Commun.}\ }\textbf {\bibinfo {volume} {6}},\ \bibinfo {pages} {8204} (\bibinfo {year} {2015})}\BibitemShut {NoStop}%
\bibitem [{\citenamefont {Somaschi}\ \emph {et~al.}(2016)\citenamefont {Somaschi}, \citenamefont {Giesz}, \citenamefont {De~Santis}, \citenamefont {Loredo}, \citenamefont {Almeida}, \citenamefont {Hornecker}, \citenamefont {Portalupi}, \citenamefont {Grange}, \citenamefont {Ant{\'o}n}, \citenamefont {Demory}, \citenamefont {G{\'o}mez}, \citenamefont {Sagnes}, \citenamefont {Lema{\^i}tre}, \citenamefont {Auffeves}, \citenamefont {White}, \citenamefont {Lanco},\ and\ \citenamefont {Senellart}}]{Somaschi2016}%
  \BibitemOpen
  \bibfield  {author} {\bibinfo {author} {\bibfnamefont {N.}~\bibnamefont {Somaschi}}, \bibinfo {author} {\bibfnamefont {V.}~\bibnamefont {Giesz}}, \bibinfo {author} {\bibfnamefont {L.}~\bibnamefont {De~Santis}}, \bibinfo {author} {\bibfnamefont {J.~C.}\ \bibnamefont {Loredo}}, \bibinfo {author} {\bibfnamefont {M.~P.}\ \bibnamefont {Almeida}}, \bibinfo {author} {\bibfnamefont {G.}~\bibnamefont {Hornecker}}, \bibinfo {author} {\bibfnamefont {S.~L.}\ \bibnamefont {Portalupi}}, \bibinfo {author} {\bibfnamefont {T.}~\bibnamefont {Grange}}, \bibinfo {author} {\bibfnamefont {C.}~\bibnamefont {Ant{\'o}n}}, \bibinfo {author} {\bibfnamefont {J.}~\bibnamefont {Demory}}, \bibinfo {author} {\bibfnamefont {C.}~\bibnamefont {G{\'o}mez}}, \bibinfo {author} {\bibfnamefont {I.}~\bibnamefont {Sagnes}}, \bibinfo {author} {\bibfnamefont {A.}~\bibnamefont {Lema{\^i}tre}}, \bibinfo {author} {\bibfnamefont {A.}~\bibnamefont {Auffeves}}, \bibinfo {author} {\bibfnamefont {A.~G.}\ \bibnamefont {White}}, \bibinfo {author}
  {\bibfnamefont {L.}~\bibnamefont {Lanco}},\ and\ \bibinfo {author} {\bibfnamefont {P.}~\bibnamefont {Senellart}},\ }\bibfield  {title} {\bibinfo {title} {Near-optimal single-photon sources in the solid state},\ }\href {https://doi.org/10.1038/nphoton.2016.23} {\bibfield  {journal} {\bibinfo  {journal} {Nat. Photonics}\ }\textbf {\bibinfo {volume} {10}},\ \bibinfo {pages} {340} (\bibinfo {year} {2016})}\BibitemShut {NoStop}%
\bibitem [{\citenamefont {Tomm}\ \emph {et~al.}(2021)\citenamefont {Tomm}, \citenamefont {Javadi}, \citenamefont {Antoniadis}, \citenamefont {Najer}, \citenamefont {L{\"o}bl}, \citenamefont {Korsch}, \citenamefont {Schott}, \citenamefont {Valentin}, \citenamefont {Wieck}, \citenamefont {Ludwig},\ and\ \citenamefont {Warburton}}]{Tomm2021}%
  \BibitemOpen
  \bibfield  {author} {\bibinfo {author} {\bibfnamefont {N.}~\bibnamefont {Tomm}}, \bibinfo {author} {\bibfnamefont {A.}~\bibnamefont {Javadi}}, \bibinfo {author} {\bibfnamefont {N.~O.}\ \bibnamefont {Antoniadis}}, \bibinfo {author} {\bibfnamefont {D.}~\bibnamefont {Najer}}, \bibinfo {author} {\bibfnamefont {M.~C.}\ \bibnamefont {L{\"o}bl}}, \bibinfo {author} {\bibfnamefont {A.~R.}\ \bibnamefont {Korsch}}, \bibinfo {author} {\bibfnamefont {R.}~\bibnamefont {Schott}}, \bibinfo {author} {\bibfnamefont {S.~R.}\ \bibnamefont {Valentin}}, \bibinfo {author} {\bibfnamefont {A.~D.}\ \bibnamefont {Wieck}}, \bibinfo {author} {\bibfnamefont {A.}~\bibnamefont {Ludwig}},\ and\ \bibinfo {author} {\bibfnamefont {R.~J.}\ \bibnamefont {Warburton}},\ }\bibfield  {title} {\bibinfo {title} {A bright and fast source of coherent single photons},\ }\href {https://doi.org/10.1038/s41565-020-00831-x} {\bibfield  {journal} {\bibinfo  {journal} {Nat. Nanotechnol.}\ }\textbf {\bibinfo {volume} {16}},\ \bibinfo {pages} {399} (\bibinfo
  {year} {2021})}\BibitemShut {NoStop}%
\bibitem [{\citenamefont {Aspuru-Guzik}\ and\ \citenamefont {Walther}(2012)}]{AspuruGuzik2012}%
  \BibitemOpen
  \bibfield  {author} {\bibinfo {author} {\bibfnamefont {A.}~\bibnamefont {Aspuru-Guzik}}\ and\ \bibinfo {author} {\bibfnamefont {P.}~\bibnamefont {Walther}},\ }\bibfield  {title} {\bibinfo {title} {Photonic quantum simulators},\ }\href {https://doi.org/10.1038/nphys2253} {\bibfield  {journal} {\bibinfo  {journal} {Nat. Phys.}\ }\textbf {\bibinfo {volume} {8}},\ \bibinfo {pages} {285} (\bibinfo {year} {2012})}\BibitemShut {NoStop}%
\bibitem [{\citenamefont {Loss}\ and\ \citenamefont {DiVincenzo}(1998)}]{Loss1998}%
  \BibitemOpen
  \bibfield  {author} {\bibinfo {author} {\bibfnamefont {D.}~\bibnamefont {Loss}}\ and\ \bibinfo {author} {\bibfnamefont {D.~P.}\ \bibnamefont {DiVincenzo}},\ }\bibfield  {title} {\bibinfo {title} {Quantum computation with quantum dots},\ }\href {https://doi.org/10.1103/PhysRevA.57.120} {\bibfield  {journal} {\bibinfo  {journal} {Phys. Rev. A}\ }\textbf {\bibinfo {volume} {57}},\ \bibinfo {pages} {120} (\bibinfo {year} {1998})}\BibitemShut {NoStop}%
\bibitem [{\citenamefont {Kuhlmann}\ \emph {et~al.}(2013)\citenamefont {Kuhlmann}, \citenamefont {Houel}, \citenamefont {Reuter}, \citenamefont {Wieck},\ and\ \citenamefont {Warburton}}]{Kuhlmann2013}%
  \BibitemOpen
  \bibfield  {author} {\bibinfo {author} {\bibfnamefont {A.~V.}\ \bibnamefont {Kuhlmann}}, \bibinfo {author} {\bibfnamefont {J.}~\bibnamefont {Houel}}, \bibinfo {author} {\bibfnamefont {D.}~\bibnamefont {Reuter}}, \bibinfo {author} {\bibfnamefont {A.~D.}\ \bibnamefont {Wieck}},\ and\ \bibinfo {author} {\bibfnamefont {R.~J.}\ \bibnamefont {Warburton}},\ }\bibfield  {title} {\bibinfo {title} {Dark-field microscopy for background-free detection of resonance fluorescence from a single semiconductor quantum dot},\ }\href {https://doi.org/10.1063/1.4813879} {\bibfield  {journal} {\bibinfo  {journal} {Rev. Sci. Instrum.}\ }\textbf {\bibinfo {volume} {84}},\ \bibinfo {pages} {073905} (\bibinfo {year} {2013})}\BibitemShut {NoStop}%
\bibitem [{\citenamefont {Ramsay}\ \emph {et~al.}(2010)\citenamefont {Ramsay}, \citenamefont {Gopal}, \citenamefont {Gauger}, \citenamefont {Nazir}, \citenamefont {Lovett}, \citenamefont {Fox},\ and\ \citenamefont {Skolnick}}]{Ramsay2010phonons}%
  \BibitemOpen
  \bibfield  {author} {\bibinfo {author} {\bibfnamefont {A.~J.}\ \bibnamefont {Ramsay}}, \bibinfo {author} {\bibfnamefont {A.~V.}\ \bibnamefont {Gopal}}, \bibinfo {author} {\bibfnamefont {E.~M.}\ \bibnamefont {Gauger}}, \bibinfo {author} {\bibfnamefont {A.}~\bibnamefont {Nazir}}, \bibinfo {author} {\bibfnamefont {B.~W.}\ \bibnamefont {Lovett}}, \bibinfo {author} {\bibfnamefont {A.~M.}\ \bibnamefont {Fox}},\ and\ \bibinfo {author} {\bibfnamefont {M.~S.}\ \bibnamefont {Skolnick}},\ }\bibfield  {title} {\bibinfo {title} {Phonon-induced dephasing of optically driven excitons in quantum dots},\ }\href {https://doi.org/10.1103/PhysRevLett.104.017402} {\bibfield  {journal} {\bibinfo  {journal} {Phys. Rev. Lett.}\ }\textbf {\bibinfo {volume} {104}},\ \bibinfo {pages} {017402} (\bibinfo {year} {2010})}\BibitemShut {NoStop}%
\bibitem [{\citenamefont {Iles-Smith}\ \emph {et~al.}(2017)\citenamefont {Iles-Smith}, \citenamefont {McCutcheon}, \citenamefont {M{\o}rk},\ and\ \citenamefont {Nazir}}]{IlesSmith2017}%
  \BibitemOpen
  \bibfield  {author} {\bibinfo {author} {\bibfnamefont {J.}~\bibnamefont {Iles-Smith}}, \bibinfo {author} {\bibfnamefont {D.~P.~S.}\ \bibnamefont {McCutcheon}}, \bibinfo {author} {\bibfnamefont {J.}~\bibnamefont {M{\o}rk}},\ and\ \bibinfo {author} {\bibfnamefont {A.}~\bibnamefont {Nazir}},\ }\bibfield  {title} {\bibinfo {title} {Phonon-assisted population inversion of a quantum dot: A theoretical and experimental investigation},\ }\href {https://doi.org/10.1103/PhysRevB.95.201305} {\bibfield  {journal} {\bibinfo  {journal} {Phys. Rev. B}\ }\textbf {\bibinfo {volume} {95}},\ \bibinfo {pages} {201305(R)} (\bibinfo {year} {2017})}\BibitemShut {NoStop}%
\bibitem [{\citenamefont {Vandersypen}\ and\ \citenamefont {Chuang}(2005)}]{Vandersypen2005}%
  \BibitemOpen
  \bibfield  {author} {\bibinfo {author} {\bibfnamefont {L.~M.~K.}\ \bibnamefont {Vandersypen}}\ and\ \bibinfo {author} {\bibfnamefont {I.~L.}\ \bibnamefont {Chuang}},\ }\bibfield  {title} {\bibinfo {title} {Nmr techniques for quantum control and computation},\ }\href {https://doi.org/10.1103/RevModPhys.76.1037} {\bibfield  {journal} {\bibinfo  {journal} {Rev. Mod. Phys.}\ }\textbf {\bibinfo {volume} {76}},\ \bibinfo {pages} {1037} (\bibinfo {year} {2005})}\BibitemShut {NoStop}%
\bibitem [{\citenamefont {Allen}\ and\ \citenamefont {Eberly}(1987)}]{Allen1987}%
  \BibitemOpen
  \bibfield  {author} {\bibinfo {author} {\bibfnamefont {L.}~\bibnamefont {Allen}}\ and\ \bibinfo {author} {\bibfnamefont {J.~H.}\ \bibnamefont {Eberly}},\ }\href@noop {} {\emph {\bibinfo {title} {Optical Resonance and Two-Level Atoms}}}\ (\bibinfo  {publisher} {Dover Publications},\ \bibinfo {address} {New York},\ \bibinfo {year} {1987})\BibitemShut {NoStop}%
\bibitem [{\citenamefont {Press}\ \emph {et~al.}(2008)\citenamefont {Press}, \citenamefont {Ladd}, \citenamefont {Zhang},\ and\ \citenamefont {Yamamoto}}]{Press2008}%
  \BibitemOpen
  \bibfield  {author} {\bibinfo {author} {\bibfnamefont {D.}~\bibnamefont {Press}}, \bibinfo {author} {\bibfnamefont {T.~D.}\ \bibnamefont {Ladd}}, \bibinfo {author} {\bibfnamefont {B.}~\bibnamefont {Zhang}},\ and\ \bibinfo {author} {\bibfnamefont {Y.}~\bibnamefont {Yamamoto}},\ }\bibfield  {title} {\bibinfo {title} {Complete quantum control of a single quantum dot spin using ultrafast optical pulses},\ }\href {https://doi.org/10.1038/nature07530} {\bibfield  {journal} {\bibinfo  {journal} {Nature}\ }\textbf {\bibinfo {volume} {456}},\ \bibinfo {pages} {218} (\bibinfo {year} {2008})}\BibitemShut {NoStop}%
\bibitem [{\citenamefont {Malinovsky}\ and\ \citenamefont {Krause}(2001)}]{Malinovsky2001}%
  \BibitemOpen
  \bibfield  {author} {\bibinfo {author} {\bibfnamefont {V.~S.}\ \bibnamefont {Malinovsky}}\ and\ \bibinfo {author} {\bibfnamefont {J.~L.}\ \bibnamefont {Krause}},\ }\bibfield  {title} {\bibinfo {title} {General theory of robust population transfer in a three-level system by suitably shaped laser pulses},\ }\href {https://doi.org/10.1103/PhysRevA.63.043415} {\bibfield  {journal} {\bibinfo  {journal} {Phys. Rev. A}\ }\textbf {\bibinfo {volume} {63}},\ \bibinfo {pages} {043415} (\bibinfo {year} {2001})}\BibitemShut {NoStop}%
\bibitem [{\citenamefont {Vitanov}\ \emph {et~al.}(2017)\citenamefont {Vitanov}, \citenamefont {Rangelov}, \citenamefont {Shore},\ and\ \citenamefont {Bergmann}}]{Vitanov2017}%
  \BibitemOpen
  \bibfield  {author} {\bibinfo {author} {\bibfnamefont {N.~V.}\ \bibnamefont {Vitanov}}, \bibinfo {author} {\bibfnamefont {A.~A.}\ \bibnamefont {Rangelov}}, \bibinfo {author} {\bibfnamefont {B.~W.}\ \bibnamefont {Shore}},\ and\ \bibinfo {author} {\bibfnamefont {K.}~\bibnamefont {Bergmann}},\ }\bibfield  {title} {\bibinfo {title} {Stimulated raman adiabatic passage in physics, chemistry, and beyond},\ }\href {https://doi.org/10.1103/RevModPhys.89.015006} {\bibfield  {journal} {\bibinfo  {journal} {Rev. Mod. Phys.}\ }\textbf {\bibinfo {volume} {89}},\ \bibinfo {pages} {015006} (\bibinfo {year} {2017})}\BibitemShut {NoStop}%
\bibitem [{\citenamefont {Wu}\ \emph {et~al.}(2011)\citenamefont {Wu}, \citenamefont {Piper}, \citenamefont {Ediger}, \citenamefont {Brereton}, \citenamefont {Schmidgall}, \citenamefont {Eastham}, \citenamefont {Hugues}, \citenamefont {Hopkinson},\ and\ \citenamefont {Phillips}}]{Wu2011}%
  \BibitemOpen
  \bibfield  {author} {\bibinfo {author} {\bibfnamefont {Y.}~\bibnamefont {Wu}}, \bibinfo {author} {\bibfnamefont {I.~M.}\ \bibnamefont {Piper}}, \bibinfo {author} {\bibfnamefont {M.}~\bibnamefont {Ediger}}, \bibinfo {author} {\bibfnamefont {P.}~\bibnamefont {Brereton}}, \bibinfo {author} {\bibfnamefont {E.~R.}\ \bibnamefont {Schmidgall}}, \bibinfo {author} {\bibfnamefont {P.~R.}\ \bibnamefont {Eastham}}, \bibinfo {author} {\bibfnamefont {M.}~\bibnamefont {Hugues}}, \bibinfo {author} {\bibfnamefont {M.}~\bibnamefont {Hopkinson}},\ and\ \bibinfo {author} {\bibfnamefont {R.~T.}\ \bibnamefont {Phillips}},\ }\bibfield  {title} {\bibinfo {title} {Population inversion in a single semiconductor quantum dot using picosecond optical pulses},\ }\href {https://doi.org/10.1103/PhysRevLett.106.067401} {\bibfield  {journal} {\bibinfo  {journal} {Phys. Rev. Lett.}\ }\textbf {\bibinfo {volume} {106}},\ \bibinfo {pages} {067401} (\bibinfo {year} {2011})}\BibitemShut {NoStop}%
\bibitem [{\citenamefont {Wei}\ \emph {et~al.}(2014)\citenamefont {Wei}, \citenamefont {He}, \citenamefont {He}, \citenamefont {Lu}, \citenamefont {Pan}, \citenamefont {Schneider}, \citenamefont {Kamp}, \citenamefont {H{\"o}fling}, \citenamefont {Reuter},\ and\ \citenamefont {Wieck}}]{Wei2014}%
  \BibitemOpen
  \bibfield  {author} {\bibinfo {author} {\bibfnamefont {Y.-J.}\ \bibnamefont {Wei}}, \bibinfo {author} {\bibfnamefont {Y.-M.}\ \bibnamefont {He}}, \bibinfo {author} {\bibfnamefont {Y.}~\bibnamefont {He}}, \bibinfo {author} {\bibfnamefont {C.-Y.}\ \bibnamefont {Lu}}, \bibinfo {author} {\bibfnamefont {J.-W.}\ \bibnamefont {Pan}}, \bibinfo {author} {\bibfnamefont {C.}~\bibnamefont {Schneider}}, \bibinfo {author} {\bibfnamefont {M.}~\bibnamefont {Kamp}}, \bibinfo {author} {\bibfnamefont {S.}~\bibnamefont {H{\"o}fling}}, \bibinfo {author} {\bibfnamefont {D.}~\bibnamefont {Reuter}},\ and\ \bibinfo {author} {\bibfnamefont {A.~D.}\ \bibnamefont {Wieck}},\ }\bibfield  {title} {\bibinfo {title} {Deterministic and robust generation of single photons from a single quantum dot with 99.5\% indistinguishability using adiabatic rapid passage},\ }\href {https://doi.org/10.1021/nl503081n} {\bibfield  {journal} {\bibinfo  {journal} {Nano Lett.}\ }\textbf {\bibinfo {volume} {14}},\ \bibinfo {pages} {6515} (\bibinfo {year}
  {2014})}\BibitemShut {NoStop}%
\bibitem [{\citenamefont {Simon}\ \emph {et~al.}(2011)\citenamefont {Simon}, \citenamefont {Belhadj}, \citenamefont {Chatel}, \citenamefont {Amand}, \citenamefont {Renucci}, \citenamefont {Lema{\^i}tre}, \citenamefont {Krebs}, \citenamefont {Dalgarno}, \citenamefont {Warburton}, \citenamefont {Marie},\ and\ \citenamefont {Urbaszek}}]{Simon2011}%
  \BibitemOpen
  \bibfield  {author} {\bibinfo {author} {\bibfnamefont {C.-M.}\ \bibnamefont {Simon}}, \bibinfo {author} {\bibfnamefont {T.}~\bibnamefont {Belhadj}}, \bibinfo {author} {\bibfnamefont {B.}~\bibnamefont {Chatel}}, \bibinfo {author} {\bibfnamefont {T.}~\bibnamefont {Amand}}, \bibinfo {author} {\bibfnamefont {P.}~\bibnamefont {Renucci}}, \bibinfo {author} {\bibfnamefont {A.}~\bibnamefont {Lema{\^i}tre}}, \bibinfo {author} {\bibfnamefont {O.}~\bibnamefont {Krebs}}, \bibinfo {author} {\bibfnamefont {P.~A.}\ \bibnamefont {Dalgarno}}, \bibinfo {author} {\bibfnamefont {R.~J.}\ \bibnamefont {Warburton}}, \bibinfo {author} {\bibfnamefont {X.}~\bibnamefont {Marie}},\ and\ \bibinfo {author} {\bibfnamefont {B.}~\bibnamefont {Urbaszek}},\ }\bibfield  {title} {\bibinfo {title} {Robust quantum dot exciton generation via adiabatic passage},\ }\href {https://doi.org/10.1103/PhysRevLett.106.166801} {\bibfield  {journal} {\bibinfo  {journal} {Phys. Rev. Lett.}\ }\textbf {\bibinfo {volume} {106}},\ \bibinfo {pages} {166801}
  (\bibinfo {year} {2011})}\BibitemShut {NoStop}%
\bibitem [{\citenamefont {Scully}\ and\ \citenamefont {Zubairy}(1997)}]{Scully1997}%
  \BibitemOpen
  \bibfield  {author} {\bibinfo {author} {\bibfnamefont {M.~O.}\ \bibnamefont {Scully}}\ and\ \bibinfo {author} {\bibfnamefont {M.~S.}\ \bibnamefont {Zubairy}},\ }\href@noop {} {\emph {\bibinfo {title} {Quantum Optics}}}\ (\bibinfo  {publisher} {Cambridge University Press},\ \bibinfo {address} {Cambridge},\ \bibinfo {year} {1997})\BibitemShut {NoStop}%
\bibitem [{\citenamefont {Breuer}\ and\ \citenamefont {Petruccione}(2002)}]{Breuer2002}%
  \BibitemOpen
  \bibfield  {author} {\bibinfo {author} {\bibfnamefont {H.-P.}\ \bibnamefont {Breuer}}\ and\ \bibinfo {author} {\bibfnamefont {F.}~\bibnamefont {Petruccione}},\ }\href@noop {} {\emph {\bibinfo {title} {The Theory of Open Quantum Systems}}}\ (\bibinfo  {publisher} {Oxford University Press},\ \bibinfo {address} {Oxford},\ \bibinfo {year} {2002})\BibitemShut {NoStop}%
\bibitem [{\citenamefont {Mollow}(1969)}]{Mollow1969}%
  \BibitemOpen
  \bibfield  {author} {\bibinfo {author} {\bibfnamefont {B.~R.}\ \bibnamefont {Mollow}},\ }\bibfield  {title} {\bibinfo {title} {Power spectrum of light scattered by two-level systems},\ }\href {https://doi.org/10.1103/PhysRev.188.1969} {\bibfield  {journal} {\bibinfo  {journal} {Phys. Rev.}\ }\textbf {\bibinfo {volume} {188}},\ \bibinfo {pages} {1969} (\bibinfo {year} {1969})}\BibitemShut {NoStop}%
\bibitem [{\citenamefont {Carmichael}(1999)}]{Carmichael1999}%
  \BibitemOpen
  \bibfield  {author} {\bibinfo {author} {\bibfnamefont {H.~J.}\ \bibnamefont {Carmichael}},\ }\href@noop {} {\emph {\bibinfo {title} {Statistical Methods in Quantum Optics 1: Master Equations and Fokker-Planck Equations}}}\ (\bibinfo  {publisher} {Springer},\ \bibinfo {address} {Berlin, Heidelberg},\ \bibinfo {year} {1999})\BibitemShut {NoStop}%
\bibitem [{\citenamefont {Krummheuer}\ \emph {et~al.}(2002)\citenamefont {Krummheuer}, \citenamefont {Axt},\ and\ \citenamefont {Kuhn}}]{Krummheuer2002}%
  \BibitemOpen
  \bibfield  {author} {\bibinfo {author} {\bibfnamefont {B.}~\bibnamefont {Krummheuer}}, \bibinfo {author} {\bibfnamefont {V.~M.}\ \bibnamefont {Axt}},\ and\ \bibinfo {author} {\bibfnamefont {T.}~\bibnamefont {Kuhn}},\ }\bibfield  {title} {\bibinfo {title} {Theory of pure dephasing and the resulting absorption line shape in semiconductor quantum dots},\ }\href {https://doi.org/10.1103/PhysRevB.65.195313} {\bibfield  {journal} {\bibinfo  {journal} {Phys. Rev. B}\ }\textbf {\bibinfo {volume} {65}},\ \bibinfo {pages} {195313} (\bibinfo {year} {2002})}\BibitemShut {NoStop}%
\bibitem [{\citenamefont {Manzano}(2020)}]{Manzano2020}%
  \BibitemOpen
  \bibfield  {author} {\bibinfo {author} {\bibfnamefont {D.}~\bibnamefont {Manzano}},\ }\bibfield  {title} {\bibinfo {title} {A short introduction to the lindblad master equation},\ }\href {https://doi.org/10.1063/1.5115323} {\bibfield  {journal} {\bibinfo  {journal} {AIP Adv.}\ }\textbf {\bibinfo {volume} {10}},\ \bibinfo {pages} {025106} (\bibinfo {year} {2020})}\BibitemShut {NoStop}%
\bibitem [{\citenamefont {Borri}\ \emph {et~al.}(2001)\citenamefont {Borri}, \citenamefont {Langbein}, \citenamefont {Schneider}, \citenamefont {Woggon}, \citenamefont {Sellin}, \citenamefont {Ouyang},\ and\ \citenamefont {Bimberg}}]{Borri2001}%
  \BibitemOpen
  \bibfield  {author} {\bibinfo {author} {\bibfnamefont {P.}~\bibnamefont {Borri}}, \bibinfo {author} {\bibfnamefont {W.}~\bibnamefont {Langbein}}, \bibinfo {author} {\bibfnamefont {S.}~\bibnamefont {Schneider}}, \bibinfo {author} {\bibfnamefont {U.}~\bibnamefont {Woggon}}, \bibinfo {author} {\bibfnamefont {R.~L.}\ \bibnamefont {Sellin}}, \bibinfo {author} {\bibfnamefont {D.}~\bibnamefont {Ouyang}},\ and\ \bibinfo {author} {\bibfnamefont {D.}~\bibnamefont {Bimberg}},\ }\bibfield  {title} {\bibinfo {title} {Ultralong dephasing time in ingaas quantum dots},\ }\href {https://doi.org/10.1103/PhysRevLett.87.157401} {\bibfield  {journal} {\bibinfo  {journal} {Physical Review Letters}\ }\textbf {\bibinfo {volume} {87}},\ \bibinfo {pages} {157401} (\bibinfo {year} {2001})}\BibitemShut {NoStop}%
\bibitem [{\citenamefont {Shore}(2011)}]{Shore2011}%
  \BibitemOpen
  \bibfield  {author} {\bibinfo {author} {\bibfnamefont {B.~W.}\ \bibnamefont {Shore}},\ }\href {https://doi.org/10.1017/CBO9780511719938} {\emph {\bibinfo {title} {Manipulating Quantum Structures Using Laser Pulses}}}\ (\bibinfo  {publisher} {Cambridge University Press},\ \bibinfo {address} {Cambridge},\ \bibinfo {year} {2011})\BibitemShut {NoStop}%
\bibitem [{\citenamefont {Ramsay}(2010)}]{Ramsay2010}%
  \BibitemOpen
  \bibfield  {author} {\bibinfo {author} {\bibfnamefont {A.~J.}\ \bibnamefont {Ramsay}},\ }\bibfield  {title} {\bibinfo {title} {Coherent optical control of a single quantum dot},\ }\href {https://doi.org/10.1088/0268-1242/25/10/103001} {\bibfield  {journal} {\bibinfo  {journal} {Semicond. Sci. Technol.}\ }\textbf {\bibinfo {volume} {25}},\ \bibinfo {pages} {103001} (\bibinfo {year} {2010})}\BibitemShut {NoStop}%
\bibitem [{\citenamefont {Stark}\ \emph {et~al.}(2022)\citenamefont {Stark}, \citenamefont {Wilbur}, \citenamefont {Meesala}, \citenamefont {Son}, \citenamefont {Allen}, \citenamefont {Lu}, \citenamefont {Bersin}, \citenamefont {Mouradian},\ and\ \citenamefont {Englund}}]{Stark2022}%
  \BibitemOpen
  \bibfield  {author} {\bibinfo {author} {\bibfnamefont {C.~J.}\ \bibnamefont {Stark}}, \bibinfo {author} {\bibfnamefont {C.~C.}\ \bibnamefont {Wilbur}}, \bibinfo {author} {\bibfnamefont {S.}~\bibnamefont {Meesala}}, \bibinfo {author} {\bibfnamefont {G.}~\bibnamefont {Son}}, \bibinfo {author} {\bibfnamefont {J.~P.}\ \bibnamefont {Allen}}, \bibinfo {author} {\bibfnamefont {T.-H.}\ \bibnamefont {Lu}}, \bibinfo {author} {\bibfnamefont {E.}~\bibnamefont {Bersin}}, \bibinfo {author} {\bibfnamefont {S.~L.}\ \bibnamefont {Mouradian}},\ and\ \bibinfo {author} {\bibfnamefont {D.}~\bibnamefont {Englund}},\ }\bibfield  {title} {\bibinfo {title} {Theory of notch-filtered adiabatic rapid passage},\ }\href {https://doi.org/10.1103/PhysRevA.105.053703} {\bibfield  {journal} {\bibinfo  {journal} {Phys. Rev. A}\ }\textbf {\bibinfo {volume} {105}},\ \bibinfo {pages} {053703} (\bibinfo {year} {2022})}\BibitemShut {NoStop}%
\bibitem [{\citenamefont {Landau}(1932)}]{Landau1932}%
  \BibitemOpen
  \bibfield  {author} {\bibinfo {author} {\bibfnamefont {L.}~\bibnamefont {Landau}},\ }\bibfield  {title} {\bibinfo {title} {Zur theorie der energieubertragung. ii},\ }\href@noop {} {\bibfield  {journal} {\bibinfo  {journal} {Phys. Z. Sowjetunion}\ }\textbf {\bibinfo {volume} {2}},\ \bibinfo {pages} {46} (\bibinfo {year} {1932})}\BibitemShut {NoStop}%
\bibitem [{\citenamefont {Zener}(1932)}]{Zener1932}%
  \BibitemOpen
  \bibfield  {author} {\bibinfo {author} {\bibfnamefont {C.}~\bibnamefont {Zener}},\ }\bibfield  {title} {\bibinfo {title} {Non-adiabatic crossing of energy levels},\ }\href {https://doi.org/10.1098/rspa.1932.0165} {\bibfield  {journal} {\bibinfo  {journal} {Proc. R. Soc. Lond. A}\ }\textbf {\bibinfo {volume} {137}},\ \bibinfo {pages} {696} (\bibinfo {year} {1932})}\BibitemShut {NoStop}%
\bibitem [{\citenamefont {Shevchenko}\ \emph {et~al.}(2010)\citenamefont {Shevchenko}, \citenamefont {Ashhab},\ and\ \citenamefont {Nori}}]{Shevchenko2010}%
  \BibitemOpen
  \bibfield  {author} {\bibinfo {author} {\bibfnamefont {S.~N.}\ \bibnamefont {Shevchenko}}, \bibinfo {author} {\bibfnamefont {S.}~\bibnamefont {Ashhab}},\ and\ \bibinfo {author} {\bibfnamefont {F.}~\bibnamefont {Nori}},\ }\bibfield  {title} {\bibinfo {title} {Landau-zener-st{\"u}ckelberg-majorana transitions in qubits and nanomechanical systems},\ }\href {https://doi.org/10.1016/j.physrep.2010.03.002} {\bibfield  {journal} {\bibinfo  {journal} {Phys. Rep.}\ }\textbf {\bibinfo {volume} {492}},\ \bibinfo {pages} {1} (\bibinfo {year} {2010})}\BibitemShut {NoStop}%
\bibitem [{\citenamefont {Eberly}\ \emph {et~al.}(1980)\citenamefont {Eberly}, \citenamefont {Kunasz},\ and\ \citenamefont {W{\'o}dkiewicz}}]{Eberly1977}%
  \BibitemOpen
  \bibfield  {author} {\bibinfo {author} {\bibfnamefont {J.~H.}\ \bibnamefont {Eberly}}, \bibinfo {author} {\bibfnamefont {C.~V.}\ \bibnamefont {Kunasz}},\ and\ \bibinfo {author} {\bibfnamefont {K.}~\bibnamefont {W{\'o}dkiewicz}},\ }\bibfield  {title} {\bibinfo {title} {The time-dependent physical spectrum of resonance fluorescence},\ }\href {https://doi.org/10.1088/0022-3700/13/2/010} {\bibfield  {journal} {\bibinfo  {journal} {J. Phys. B: At. Mol. Phys.}\ }\textbf {\bibinfo {volume} {13}},\ \bibinfo {pages} {217} (\bibinfo {year} {1980})}\BibitemShut {NoStop}%
\bibitem [{\citenamefont {Lax}(1963)}]{Lax1963}%
  \BibitemOpen
  \bibfield  {author} {\bibinfo {author} {\bibfnamefont {M.}~\bibnamefont {Lax}},\ }\bibfield  {title} {\bibinfo {title} {Formal theory of quantum fluctuations from a driven state},\ }\href {https://doi.org/10.1103/PhysRev.129.2342} {\bibfield  {journal} {\bibinfo  {journal} {Phys. Rev.}\ }\textbf {\bibinfo {volume} {129}},\ \bibinfo {pages} {2342} (\bibinfo {year} {1963})}\BibitemShut {NoStop}%
\bibitem [{\citenamefont {Swain}(1981)}]{Swain1981}%
  \BibitemOpen
  \bibfield  {author} {\bibinfo {author} {\bibfnamefont {S.}~\bibnamefont {Swain}},\ }\bibfield  {title} {\bibinfo {title} {A model for resonance fluorescence},\ }\href {https://doi.org/10.1088/0022-3700/14/15/014} {\bibfield  {journal} {\bibinfo  {journal} {J. Phys. B: At. Mol. Phys.}\ }\textbf {\bibinfo {volume} {14}},\ \bibinfo {pages} {2577} (\bibinfo {year} {1981})}\BibitemShut {NoStop}%
\bibitem [{\citenamefont {Claudon}\ \emph {et~al.}(2010)\citenamefont {Claudon}, \citenamefont {Bleuse}, \citenamefont {Malik}, \citenamefont {Bazin}, \citenamefont {Jaffrennou}, \citenamefont {Gregersen}, \citenamefont {Sauvan}, \citenamefont {Lalanne},\ and\ \citenamefont {G{\'e}rard}}]{Claudon2010}%
  \BibitemOpen
  \bibfield  {author} {\bibinfo {author} {\bibfnamefont {J.}~\bibnamefont {Claudon}}, \bibinfo {author} {\bibfnamefont {J.}~\bibnamefont {Bleuse}}, \bibinfo {author} {\bibfnamefont {N.~S.}\ \bibnamefont {Malik}}, \bibinfo {author} {\bibfnamefont {M.}~\bibnamefont {Bazin}}, \bibinfo {author} {\bibfnamefont {P.}~\bibnamefont {Jaffrennou}}, \bibinfo {author} {\bibfnamefont {N.}~\bibnamefont {Gregersen}}, \bibinfo {author} {\bibfnamefont {C.}~\bibnamefont {Sauvan}}, \bibinfo {author} {\bibfnamefont {P.}~\bibnamefont {Lalanne}},\ and\ \bibinfo {author} {\bibfnamefont {J.-M.}\ \bibnamefont {G{\'e}rard}},\ }\bibfield  {title} {\bibinfo {title} {A highly efficient single-photon source based on a quantum dot in a photonic nanowire},\ }\href {https://doi.org/10.1038/nphoton.2009.287} {\bibfield  {journal} {\bibinfo  {journal} {Nat. Photonics}\ }\textbf {\bibinfo {volume} {4}},\ \bibinfo {pages} {174} (\bibinfo {year} {2010})}\BibitemShut {NoStop}%
\bibitem [{\citenamefont {Arcari}\ \emph {et~al.}(2014)\citenamefont {Arcari}, \citenamefont {S{\"o}llner}, \citenamefont {Javadi}, \citenamefont {Lindskov~Hansen}, \citenamefont {Mahmoodian}, \citenamefont {Liu}, \citenamefont {Thyrrestrup}, \citenamefont {Lee}, \citenamefont {Song}, \citenamefont {Stobbe},\ and\ \citenamefont {Lodahl}}]{Arcari2014}%
  \BibitemOpen
  \bibfield  {author} {\bibinfo {author} {\bibfnamefont {M.}~\bibnamefont {Arcari}}, \bibinfo {author} {\bibfnamefont {I.}~\bibnamefont {S{\"o}llner}}, \bibinfo {author} {\bibfnamefont {A.}~\bibnamefont {Javadi}}, \bibinfo {author} {\bibfnamefont {S.}~\bibnamefont {Lindskov~Hansen}}, \bibinfo {author} {\bibfnamefont {S.}~\bibnamefont {Mahmoodian}}, \bibinfo {author} {\bibfnamefont {J.}~\bibnamefont {Liu}}, \bibinfo {author} {\bibfnamefont {H.}~\bibnamefont {Thyrrestrup}}, \bibinfo {author} {\bibfnamefont {E.~H.}\ \bibnamefont {Lee}}, \bibinfo {author} {\bibfnamefont {J.~D.}\ \bibnamefont {Song}}, \bibinfo {author} {\bibfnamefont {S.}~\bibnamefont {Stobbe}},\ and\ \bibinfo {author} {\bibfnamefont {P.}~\bibnamefont {Lodahl}},\ }\bibfield  {title} {\bibinfo {title} {Near-unity coupling efficiency of a quantum emitter to a photonic crystal waveguide},\ }\href {https://doi.org/10.1103/PhysRevLett.113.093603} {\bibfield  {journal} {\bibinfo  {journal} {Phys. Rev. Lett.}\ }\textbf {\bibinfo {volume} {113}},\
  \bibinfo {pages} {093603} (\bibinfo {year} {2014})}\BibitemShut {NoStop}%
\bibitem [{\citenamefont {Vagov}\ \emph {et~al.}(2007)\citenamefont {Vagov}, \citenamefont {Axt}, \citenamefont {Kuhn}, \citenamefont {Langbein}, \citenamefont {Borri},\ and\ \citenamefont {Woggon}}]{Vagov2007}%
  \BibitemOpen
  \bibfield  {author} {\bibinfo {author} {\bibfnamefont {A.}~\bibnamefont {Vagov}}, \bibinfo {author} {\bibfnamefont {V.~M.}\ \bibnamefont {Axt}}, \bibinfo {author} {\bibfnamefont {T.}~\bibnamefont {Kuhn}}, \bibinfo {author} {\bibfnamefont {W.}~\bibnamefont {Langbein}}, \bibinfo {author} {\bibfnamefont {P.}~\bibnamefont {Borri}},\ and\ \bibinfo {author} {\bibfnamefont {U.}~\bibnamefont {Woggon}},\ }\bibfield  {title} {\bibinfo {title} {Nonmonotonous temperature dependence of the zero-phonon line in quantum dots},\ }\href {https://doi.org/10.1103/PhysRevB.75.115332} {\bibfield  {journal} {\bibinfo  {journal} {Phys. Rev. B}\ }\textbf {\bibinfo {volume} {75}},\ \bibinfo {pages} {115332} (\bibinfo {year} {2007})}\BibitemShut {NoStop}%
\end{thebibliography}%


\end{document}